\documentclass[letterpaper,10pt,twocolumn,twoside,journal]{IEEEtran}

\usepackage{lmodern}

\usepackage{cite}

\usepackage{amsfonts,amssymb,amsmath} 
\usepackage{graphicx,graphics,epsfig,color}
\usepackage{preamble/rMacros}
\usepackage{hyperref}

\usepackage{enumitem}
\usepackage{balance}
\usepackage{stackengine}

\definecolor{olivegreen}{rgb}{0.14,0.29,0}

\newif\ifitsdraft

\newenvironment{proof}{{\it Proof. }}{\hfill $\blacksquare$ } 

\newtheorem{exe}{Example}
\newtheorem{corol}{Corollary}
\newtheorem{ass}{Assumption}
\newtheorem{defin}{Definition}
\newtheorem{prob}{Problem}
\newtheorem{cla}{Claim}
\newtheorem{rem}{Remark}
\newtheorem{lem}{Lemma}
\newtheorem{prop}{Proposition}
\newtheorem{thm}{Theorem}
\newtheorem{fct}{Fact}
\newtheorem{conj}{Conjecture}

\newenvironment{proposition}{\begin{prop}}{\hfill $\square$ \end{prop}}

\newenvironment{remark}{\begin{rem} \rm}{\hfill $\blacktriangle $ \end{rem}}
\newenvironment{assumption}{\begin{ass}}{\hfill $\bullet$ \end{ass}}
\newenvironment{theorem}{\begin{thm}}{\hfill $\square$ \end{thm}}
\newenvironment{definition}{\begin{defin}}{\hfill $\bullet$ \end{defin}}

\renewcommand{\iff}{\Leftrightarrow}

\usepackage[rgb]{xcolor}
\definecolor{darkgreen}{rgb}{0.0, 0.5, 0.0}

\usepackage{balance}
\usepackage{wrapfig}
\usepackage[utf8]{inputenc}
\usepackage{chngcntr}
\usepackage{apptools}
\AtAppendix{\counterwithin{theorem}{section}}

\usepackage{tikz} 
\usepackage{mathtools}
\mathtoolsset{showonlyrefs} % To hide unsuded equation numbers
\usepackage[normalem]{ulem}
\usepackage{microtype}
\usepackage{booktabs}

{\begin{mdframed}\begin{trm}}
		{\end{trm}\end{mdframed}}

\usepackage{soul}
\usepackage{cancel}

\title{\Large \bf 
	Algebraic Characterization of Time-Varying Set Invariance} 

\author{S. B. Alaoui, E.J. Wafo Wembe , A. Saoud
	\thanks{S. B. Alaoui,  E.J. WAFO WEMBE and A. Saoud are with CLIC LAB, College of Computing, University Mohammed VI Polytechnic (UM6P), Benguerir, Morocco.  }
}

\begin{document}
	\newcommand\barbelow[1]{\stackunder[1.2pt]{$#1$}{\rule{.8ex}{.075ex}}} 
	
	\maketitle
	
\begin{abstract}
	The verification of temporal forward invariance for dynamical systems subject to time-varying constraints is a foundational problem in the analysis of safety-critical systems. Existing methodologies rely on restrictive geometric assumptions, such as the Linear Independence Constraint Qualification (LICQ), or suffer from exponential computational complexity when applied to time-varying polytopes. This paper establishes a rigorous, unified, and computationally tractable framework for temporal safety verification. First, utilizing tools from nonsmooth analysis, we prove that a Nagumo-type geometric inclusion evaluated strictly on the boundary of the augmented graph is both necessary and sufficient for temporal forward invariance. Second, we provide an algebraic characterization of this condition requiring only the weaker Abadie Constraint Qualification (ACQ), generalizing standard Control Barrier Function (CBF) boundary conditions to non-smooth, multi-constraint domains. Finally, for the practical class of Linear Time-Invariant (LTI) systems with time-varying polytopic constraints, we deploy Farkas' Lemma to eliminate exponential vertex enumeration. By introducing a Lipschitz-bounded time-discretization scheme, we reduce the infinite-dimensional continuous-time verification problem to a finite, highly tractable set of decoupled Linear Programs (LPs). 
\end{abstract}

\vspace{-5pt}
\section{Introduction}
\label{sec:introduction}

The deployment of autonomous systems in complex, dynamic environments necessitates strict mathematical guarantees of safety, typically formulated as set invariance. While classical invariance assumes static domains, modern applications---such as navigation among moving obstacles, spacecraft docking, and systems governed by Signal Temporal Logic (STL)---mandate satisfaction of \textit{time-varying} constraints, giving rise to the problem of \textit{temporal forward invariance}.
	
	The theoretical foundations of invariance are deeply rooted in viability theory and differential inclusions. The classical Nagumo theorem \cite{nagumo1942lage} provides a necessary and sufficient geometric condition for the positive invariance of a {static} closed set by analyzing the system's vector field against the contingent cone of the set. Extending these geometric constructs to time-varying domains and translating them into algorithmically verifiable conditions presents significant theoretical and computational challenges: boundary non-smoothness, multi-constraint degeneracy, and the continuous-time nature of the verification horizon each introduce distinct obstacles.
	
	Recent advances, particularly Control Barrier Functions (CBFs) \cite{ames2019control}, have popularized algebraic methods for safety enforcement. However, CBF-based methods are primarily synthesis-oriented, yielding only sufficient conditions for invariance, and critically depend on the Linear Independence Constraint Qualification (LICQ) or strict transversality conditions, both of which fail categorically at the degenerate vertices of polytopic sets. For LTI systems subject to time-varying polytopes, geometric invariance verification requires checking the system vector field at every vertex of every active face---a procedure whose complexity grows exponentially with the state dimension. Time-varying constraints compound this further by requiring verification over a continuous-time horizon, rendering the problem infinite-dimensional.
	
	To bridge the gap between abstract nonsmooth analysis and tractable computational verification, this paper makes the following principal contributions:
	\vspace{-5pt}
	\begin{enumerate}
		\item We rigorously establish a Nagumo-type {equivalence} for temporal forward invariance, proving that the geometric inclusion of the augmented vector field within the temporal contingent cone, verified {strictly on the boundary} $\partial\,\mathrm{Gr}(K)$, is both necessary and sufficient. This sharpens prior viability-theory results that require verification over the full interior of the graph.
		\item We translate this geometric condition into a verifiable algebraic form via the Abadie Constraint Qualification (ACQ). This formulation unifies and generalizes existing CBF boundary conditions to multi-constraint, non-smooth sets where LICQ fails, and recovers the classical time-varying CBF condition of \cite{ames2019control} as a special case.
		\item For LTI systems constrained by time-varying polytopes, we bypass the combinatorial explosion of vertex enumeration entirely. By deploying Farkas' Lemma, we transform the geometric containment problem into an algebraic matrix feasibility condition.
		\item We introduce a time-discretization scheme augmented with a Lipschitz-derived robustness margin, reducing the continuous-time verification continuum to a finite set of decoupled Linear Programs (LPs) with worst-case polynomial time complexity $\mathcal{O}(N \cdot m^{3.5})$.
	\end{enumerate}
	
	The remainder of this paper is organized as follows. Section~\ref{sec_preliminaries} introduces the preliminary concepts of differential inclusions and nonsmooth analysis. Section~\ref{sec_main_results} presents the main theoretical results, transitioning from geometric to algebraic characterizations. The computational framework for LTI systems and time-varying polytopes is developed in Sections~\ref{sec:problem_formulation_lti}--\ref{sec:robust_disc}, and case studies are presented in Section~\ref{sec:case_studies}.

\section{Related Work}
\label{sec:related_work}
Set invariance is foundational in dynamical systems and control \cite{blanchini2008set}, with viability theory and nonsmooth analysis providing the tools for complex domains \cite{Aubin1991Viability,clarkeOptimizationNonsmoothAnalysis1990}. Standard graph-based extensions of Nagumo's theorem are often conservative or require vertex enumeration. Instead, we use proximal normal and contingent cones to establish exact necessary and sufficient conditions directly on the moving boundary $\partial K(t)$ for domains with time-varying $H(t)$ and $h(t)$.Control Barrier Functions (CBFs) are widely used for safety synthesis \cite{ames2019control}, extending to high-relative-degree systems \cite{xiao2022sufficient}, multi-constraint compatibility \cite{tan2022compatibility}, time-varying specifications \cite{lindemann2019cbf,lindemann2020decentralized}, and robust contraction rates \cite{black2023safety}. However, CBFs are synthesis-oriented, provide only sufficient conditions, and rely on LICQ. In contrast, our framework leverages the Abadie Constraint Qualification (ACQ)—which holds naturally at non-smooth polyhedral vertices—enabling exact, vertex-free invariance verification.Reachability methods for LTI systems \cite{kurzhanski2014dynamics}, such as zonotopes and support functions \cite{althoff2021set,le2010reachability}, compute reachable sets rather than certify invariance of a prescribed time-varying polytope $K(t)$. Combining Farkas’ Lemma with localized margins $(L(t_k), \sigma(t_k))$, we reduce continuous viability to pointwise Metzler-like Linear Programs. This provides an exact invariance verification method with polynomial complexity in constraint count $m$ and time steps $N$, ensuring continuous inter-sampling safety across $[t_k, t_{k+1}]$.
	\vspace{-5pt}
	
	\section{Preliminaries}
	\label{sec_preliminaries}

	\textbf{Notation.}
	The set of real and non-negative real numbers are denoted by $\mathbb{R}$ and $\mathbb{R}_{\ge 0}$, respectively. For a set-valued map $K: \mathbb{R}_{\ge 0} \rightrightarrows \mathbb{R}^n$, $\text{Gr}(K)$ denotes its graph in the space $\mathbb{R} \times \mathbb{R}^n$. The boundary and interior of a set $S$ are denoted by $\partial S$ and $\text{int}(S)$. For a polytope $\mathcal{P}$, $\text{vert}(\mathcal{P})$ refers to the set of its extreme points (vertices). Given a vector $x$ and a closed set $S$, $\text{Proj}_S(x)$ is the set of metric projections of $x$ onto $S$, and $\|x\|_S := \inf_{y \in S} \|x - y\|$ denotes the distance from $x$ to $S$. For a differentiable function $g(x,t)$, $\nabla_x g$ represents the gradient with respect to $x$, while $\dot{H}(t)$ represents the element-wise time derivative. For a set-valued map $F$, $\text{dom}\,F$ and $\text{Im}(F)$ denote its effective domain and image, respectively. Finally, $F : \mathbb{R}^n \rightrightarrows \mathbb{R}^n$ is said to satisfy the {one-sided local Lipschitz} (OSL) property if, for each non-empty compact set $\mathcal{N} \subset \text{dom}\,F$, there exists $k>0$ such that
	\begin{equation}
		\begin{split}
			(x_1 - x_2)^\top F(x_1) \subset &(x_1 - x_2)^\top  F(x_2) + k |x_1-x_2|^2 \mathbb{B} \\ 
			& \forall (x_1,x_2) \in \mathcal{N}  \times \mathcal{N},
		\end{split}
	\end{equation} 
    where the inclusion $\subset$ is understood in the sense of the set-valued inner product, i.e., $\langle x_1-x_2, F(x_1)\rangle_{\sup} \le \langle x_1-x_2, F(x_2)\rangle_{\sup} + k|x_1-x_2|^2$, with $\langle x, S \rangle_{\sup} := \sup_{v \in S}\langle x, v\rangle$ for a set $S \subset \mathbb{R}^n$. The map $F$ is said to be {locally Lipschitz} if, for each non-empty compact $\mathcal{N} \subset \text{dom}\,F$, there exists $k>0$ such that
	\begin{equation}
		\begin{split}
			F(x_1) \subset   F(x_2) + k |x_1-x_2| \mathbb{B}, \qquad  \forall (x_1,x_2) \in \mathcal{N}  \times \mathcal{N}. 
		\end{split} \blacktriangle
	\end{equation}

	Let us consider the differential inclusion,
	\begin{equation}\label{eq:system}
		\Sigma: \quad \dot{x} \in f(x,t) \qquad x \in X \subset \mathbb{R}^n,
	\end{equation}
	where the set $X$ and the set-valued map $f : \mathbb{R}^n \times \mathbb{R}_{\geq 0} \rightrightarrows \mathbb{R}^n$,  verifies the following standing assumption.
	\begin{itemize} 
		\item [(SA)] The set $X$ is closed;  $f(x,t)$ is non-empty,  closed,  and convex for all $x \in X$,  and $f$ is continuous and verifies the one-sided local Lipschitz property jointly with respect to the augmented state $z = (t, x)$.
	\end{itemize}
	
	The reader is referred to~\cite[Definition 2.1.2]{Aubin1991Viability} for the definitions of lower and upper semicontinuity, and continuity, in the context of set-valued maps. The following definition outlines the concept of solutions for the systems in question.
	
	\begin{definition}\label{DefConsol}
		The function $\phi : \dom \phi \rightarrow \text{Im}(\phi)$ is a solution to $\Sigma$ if $\text{Im}(\phi) \subset X$,
		\begin{itemize}
			\item Either $\dom \phi = [0,T]$ for some $T \geq 0$ or $\dom \phi = [0,T)$ for some $T \in \mathbb{R}_{\geq 0} \cup \{+ \infty\}$, and 
			\item The function $\phi$ is locally absolutely continuous and $ \dot{\phi}(t) \in f(\phi(t))$ for almost all  $t \in \dom \phi$.   
		\end{itemize}
	\end{definition}
	
	Consider a set-valued map $K: \mathbb{R}_{\ge 0} \rightrightarrows \mathbb{R}^n$ satisfying the following regularity hypotheses: 
	\begin{description}
		\item[$(\star)$] for every $t \ge 0$, the set $K(t)$ is closed, and the map $t \mapsto K(t)$ is upper semi-continuous with respect to the Hausdorff metric.
	\end{description}
	
	Let $\phi(\cdot)$ be a solution to the system $\Sigma$ defined on an interval $[t_0, T)$. We say that $\phi$ leaves $K$ immediately at $t_0$ if $\phi(t_0) \in K(t_0)$ and there exists a constant $\delta > 0$ such that:$$\phi(t) \notin K(t), \quad \forall t \in (t_0, t_0 + \delta].$$
	
	Following the conventional definition of forward invariance in \cite{Aubin1991Viability}, we next define the temporal forward invariance concept.
	
	\begin{definition}[Temporal Forward invariance] 
		The time-varying set $K : \mathbb{R}_{\ge 0} \rightrightarrows \mathbb{R}^n$ is said to be temporally forward invariant for the system $\Sigma$ if, for every initial time $t_0 \in \mathbb{R}_{\ge 0}$ and every initial state $x_0 \in K(t_0)$, every solution $\phi(\cdot)$ starting at $\phi(t_0) = x_0$ satisfies:
		$$\phi(t) \in K(t), \quad \forall t \in [t_0, T),$$
		where $[t_0, T)$ is the domain of existence of the solution.
	\end{definition}
	
	Throughout this study, we extend the classical nonsmooth analysis constructs to the time-varying setting by considering the graph of the set-valued map $K: \mathbb{R}_{\ge 0} \rightrightarrows \mathbb{R}^n$, defined formally as a subset of the product space $\mathbb{R}_{\ge 0} \times \mathbb{R}^n$:
	$$\text{Gr}(K) := \left\{ (t, x) \in \mathbb{R}_{\ge 0} \times \mathbb{R}^n \mid x(t) \in K(t) \right\}.$$
	Assumption ($\star$) is sufficient to guarantee that the graph $\text{Gr}(K)$ is a closed subset of $\mathbb{R}_{\ge 0} \times \mathbb{R}^n$. This allows us to rigorously define the geometric cones at any point on the boundary without ambiguity \cite{Aubin1991Viability}.
	
	\begin{itemize}
		\item The \textit{temporal contingent cone} to the graph $\text{Gr}(K)$ at $z = (t, x) \in \text{Gr}(K)$ is defined as the set of vectors $\zeta \in \mathbb{R}^{n+1}$ such that $\liminf_{h \to 0^+} \frac{\text{dist}(z + h \zeta, \text{Gr}(K))}{h} = 0$. We define the \textit{temporal contingent cone} $T_K(t, x)$ as the spatial projection of the graph's contingent cone evaluated at unit time velocity:
		$$T_K(t, x) := \left\{ v \in \mathbb{R}^n \mid (1, v) \in T_{\text{Gr}(K)}(t, x) \right\}.$$
		
		\item The \textit{proximal normal cone} $N^P_{\text{Gr}(K)}(z)$ consists of all vectors $\xi \in \mathbb{R}^{n+1}$ such that there exists a scalar $\sigma > 0$ for which $z$ is the unique metric projection of $z + \sigma \xi$ onto $\text{Gr}(K)$. Equivalently, $\xi \in N^P_{\text{Gr}(K)}(z)$ if and only if there exists $\sigma > 0$ such that $\langle \xi, \tilde{z} - z\rangle \le \sigma\|\tilde{z}-z\|^2$ for all $\tilde{z} \in \text{Gr}(K)$ in a neighborhood of $z$ \cite{clarkeOptimizationNonsmoothAnalysis1990}. For any $z \in \text{Gr}(K)$, the polarity relation holds:
		\begin{equation} \label{eq:polarity}
			\langle \xi, \zeta \rangle \le 0 \qquad \forall \xi \in N^P_{\text{Gr}(K)}(z), \quad \forall \zeta \in T_{\text{Gr}(K)}(z).
	\end{equation}
	\end{itemize}
	
	Although  the aforementioned cones are sometimes introduced differently, the definitions we recall here are particularly useful for our proofs.
	
  \section{Main Results: }\label{sec_main_results}
	
	\subsection{Geometric characterization of temporal forward invariance}
	We are now ready to state our version of Nagumo-type Theorem for temporal forward invariance. 
	
	\begin{theorem}\label{Thm:Equivalence} Consider the system $\Sigma$ satisfying Assumption (SA) and the time-varying set $K$ satisfying Assumption ($\star$). The set $K(t)$ is temporally forward invariant for $\Sigma$ if and only if:
		\begin{equation} \label{eq:TheoremCondition}f(x, t) \subseteq T_K(t, x) \qquad \forall (t, x) \in \partial \text{Gr}(K).
		\end{equation}
	\end{theorem}
	\begin{proof}
		The proof is established by lifting the time-varying dynamics into the augmented state space, thereby converting the problem into a static invariance problem on the graph of $K$. Let $z := (t, x) \in \mathbb{R}^{n+1}$ denote the augmented state. We define the augmented set $\mathcal{S} := \text{Gr}(K) \subset \mathbb{R}^{n+1}$ and the augmented dynamics $F: \mathbb{R}^{n+1} \rightrightarrows \mathbb{R}^{n+1}$ as:
		\begin{equation}
			F(z) := \begin{pmatrix} 1 \\ f(x, t) \end{pmatrix}.
		\end{equation}
		Under Assumption ($\star$), $\mathcal{S}$ is a closed set. Under Assumption (SA), the augmented map $F$ inherits the properties of $f$. Specifically, $F(z)$ is non-empty, closed, convex, and satisfies the one-sided local Lipschitz property in the augmented space.
		We first relate the temporal contingent cone $T_K(t,x)$ to the standard contingent cone of the set $\mathcal{S}$. By definition of $T_K(t, x)$, we have the equivalence:
		\begin{equation} \label{eq:ConeEquivalence}
			v \in T_K(t, x) \iff \begin{pmatrix} 1 \\ v \end{pmatrix} \in T_{\mathcal{S}}(z).
		\end{equation}
		Consequently, the condition \eqref{eq:TheoremCondition} is equivalent to the static Nagumo condition on the augmented set:
		\begin{equation} \label{eq:AugmentedNagumo}
			F(z) \subseteq T_{\mathcal{S}}(z) \qquad \forall z \in \partial \mathcal{S}.
		\end{equation}
		\textbf{($\Rightarrow$)} Assume $K$ is temporally forward invariant. Let $(t, x) \in \partial \text{Gr}(K)$ and let $v \in f(x, t)$ be an arbitrary velocity vector. Since $\Sigma$ is a differential inclusion satisfying (SA), for almost all $t$ there exists a solution $\phi(\cdot)$ starting at $\phi(t) = x$ with $\dot{\phi}(t) = v$. By invariance, $(s, \phi(s)) \in \text{Gr}(K)$ for all $s \ge t$. Constructing the difference quotient for the augmented state $\zeta(s) = (s, \phi(s))$:
		$$\lim_{h \to 0^+} \frac{\zeta(t+h) - \zeta(t)}{h} = \begin{pmatrix} 1 \\ \dot{\phi}(t) \end{pmatrix} = \begin{pmatrix} 1 \\ v \end{pmatrix}.$$
		Since the trajectory remains in $\mathcal{S}$, this limit vector must belong to the contingent cone $T_{\mathcal{S}}(z)$ (see \cite[Theorem 4.1.2]{Aubin1991Viability}). By the cone equivalence in  \eqref{eq:ConeEquivalence}, this implies $v \in T_K(t, x)$. Since $v$ was arbitrary, $f(x, t) \subseteq T_K(t, x)$.

		\textbf{($\Leftarrow$)} Assume condition \eqref{eq:TheoremCondition} holds. Equivalently, $F(z) \subseteq T_{\mathcal{S}}(z)$ for all $z \in \partial \mathcal{S}$. To prove forward invariance, we demonstrate that the distance to the set cannot grow. Let $\zeta(s)$ be a trajectory of the augmented system starting at $\zeta(t_0) \in \mathcal{S}$. Since $\mathcal{S}$ is closed, the projection $\text{Proj}_{\mathcal{S}}(z)$ is non-empty. For points $z$ sufficiently close to $\mathcal{S}$, we select a projection $p \in \text{Proj}_{\mathcal{S}}(z)$. The vector $z - p$ is, by definition, a \textit{proximal normal} to $\mathcal{S}$ at $p$, denoted $z - p \in N^P_{\mathcal{S}}(p)$. We evaluate the Dini derivative of the squared distance $\Delta(z) = \frac{1}{2}\|z - p\|^2$ along the trajectory $\dot{z} \in F(z)$:
		$$D^+ \Delta(z) = \sup_{v \in F(z)} \langle z - p, v \rangle.$$
		We decompose the inner product by introducing the velocity at the projection point, $u \in F(p)$:
		$$\langle z - p, v \rangle = \langle z - p, v - u \rangle + \langle z - p, u \rangle.$$
		\begin{enumerate}
			\item[(i)] From the One-Sided Lipschitz property of $F$, we have $\langle z - p, v - u \rangle \le k \|z - p\|^2 = 2k \Delta(z)$.
			\item[(ii)] The condition $F(p) \subseteq T_{\mathcal{S}}(p)$ implies that the velocity $u$ is in the contingent cone. The polarity between the tangent and proximal normal cones, \cite[Definition 3.2.2]{Aubin1991Viability} implies that: $\langle n, w \rangle \le 0$ for all $n \in N^P_{\mathcal{S}}(p)$ and $w \in T_{\mathcal{S}}(p)$. Since $z - p \in N^P_{\mathcal{S}}(p)$ and $u \in T_{\mathcal{S}}(p)$, it follows that $\langle z - p, u \rangle \le 0$. 
		\end{enumerate}
		Combining these, we obtain the differential inequality valid locally in a neighborhood of $\partial \mathcal{S}$:$$D^+ \Delta(\zeta(s)) \le 2k \Delta(\zeta(s)).$$
		Since the initial state is in the set, $\Delta(\zeta(t_0)) = 0$. By Grönwall's Lemma, the explicit upper bound is:
		$$
		\Delta(\zeta(s)) \le \Delta(\zeta(t_0)) e^{2k(s - t_0)} = 0 \cdot e^{2k(s - t_0)} = 0,
		$$
		which implies $\Delta(\zeta(s)) = 0$ for all $s \ge t_0$, ensuring $\zeta(s) \in \mathcal{S}$.
	\end{proof}
	\subsection{Algebraic characterization of temporal forward invariance}
	Having established the geometric equivalence in Theorem \ref{Thm:Equivalence}, we now translate the geometric condition \eqref{eq:TheoremCondition} into an algebraic condition. We assume the time-varying set is defined by inequality constraints:
	\begin{equation} \label{eq:LevelSetDef}
		K(t) := \left\{ x \in \mathbb{R}^n \mid g_i(x, t) \le 0, \quad \forall i \in \{1, \dots, m\} \right\},
	\end{equation}
	where $g: \mathbb{R}^n \times \mathbb{R} \to \mathbb{R}^m$ is a vector of constraint functions. For any point $(t, x) \in \text{Gr}(K)$, let $I(x, t) := \{ i \in \{1, \dots, m\} \mid g_i(x, t) = 0 \}$ be the set of active constraints.
	
	To characterize the tangent cone algebraically, we first define the \textit{linearized cone}. Let $z := (t, x)$ denote the augmented state and let $\nabla_z g_i(z) = (\frac{\partial g_i}{\partial t}, \nabla_x g_i)$ denote the gradient in the augmented space. The linearized cone $L_{\text{Gr}(K)}(z)$ consists of all directions that satisfy the first-order approximation of the active constraints:
	\begin{equation}
		L_{\text{Gr}(K)}(z) := \left\{ \eta \in \mathbb{R}^{n+1} \mid \nabla_z g_i(z) \cdot \eta \le 0, \quad \forall i \in I(x, t) \right\}.
	\end{equation}
	It is worth noting that the geometric tangent cone is always contained within the linearized cone, i.e., $T_{\text{Gr}(K)}(z) \subseteq L_{\text{Gr}(K)}(z)$. However, equality does not hold in general without satisfying a constraint qualification. While the Linear Independence Constraint Qualification (LICQ) is commonly used \cite{black2023safety}, it is overly restrictive for our purposes as it fails at degenerate vertices of polytopes where more than $n$ constraints intersect. Therefore, we adopt the weaker \textit{Abadie Constraint Qualification} (ACQ).
	
	\begin{definition}
		The set $\text{Gr}(K)$ is said to satisfy the Abadie Constraint Qualification (ACQ) at a point $z \in \text{Gr}(K)$ if the geometric tangent cone coincides with the linearized cone:
		\begin{equation}
			T_{\text{Gr}(K)}(z) = L_{\text{Gr}(K)}(z).
		\end{equation}
	\end{definition}
	
	In the sequel, we assume the constraint functions satisfy the following regularity property.
	
	\begin{assumption}\label{ass:ACQ}
		The functions $g_i$ are continuously differentiable ($C^1$). Furthermore, the ACQ holds for every point $(t, x)$ on the boundary $\partial \text{Gr}(K)$.
	\end{assumption}
	
	\begin{remark}
		Assumption \ref{ass:ACQ} is the minimal requirement to equate the geometric and algebraic invariance conditions. Unlike LICQ, which requires the gradients of active constraints to be linearly independent, ACQ permits linear dependence. Crucially, it is well-established that polyhedral sets defined by linear constraints automatically satisfy ACQ everywhere, even at degenerate vertices \cite[Prop. 3.2.1]{Bertsekas1999}. Thus, for the time-varying polytopes considered later in this work, this assumption is satisfied inherently without imposing rank conditions on the constraint matrix.
	\end{remark}
	
	The temporal forward invariance condition can now be checked via the interaction between the system dynamics and the gradient of the constraints.
	
	\begin{theorem} \label{thm:AlgebraicBarrier}
		Consider the differential inclusion \eqref{eq:system} and the set $K(t)$ satisfying Assumption \ref{ass:ACQ}. The set $K(t)$ is temporally forward invariant for system \eqref{eq:system} if and only if for every $(t, x) \in \partial \text{Gr}(K)$ and for every active constraint $i \in I(x, t)$, the following condition holds:
		\begin{equation} \label{eq:C1_Condition}
			\sup_{v \in f(x, t)} \left( \frac{\partial g_i}{\partial t}(x, t) + \nabla_x g_i(x, t) \cdot v \right) \le 0.
		\end{equation}
	\end{theorem}
	
	\begin{proof}
		We proceed by linking the geometric tangent cone to the algebraic linearized cone via the ACQ assumption.
		Let $z := (t, x)$ be the augmented state and $\mathcal{S} := \text{Gr}(K)$. From Theorem \ref{Thm:Equivalence}, invariance is equivalent to the geometric inclusion:
		\begin{equation} \label{eq:GeoProof}
			\begin{pmatrix} 1 \\ v \end{pmatrix} \in T_{\mathcal{S}}(z) \quad \forall v \in f(x, t), \forall z \in \partial \mathcal{S}.
		\end{equation}
		
		By Assumption \ref{ass:ACQ}, the geometric tangent cone equals the linearized cone, i.e., $T_{\mathcal{S}}(z) = L_{\mathcal{S}}(z)$. Substituting this equality into \eqref{eq:GeoProof}, the invariance condition becomes:
		$$ \begin{pmatrix} 1 \\ v \end{pmatrix} \in L_{\mathcal{S}}(z) \iff \nabla_z g_i(z) \cdot \begin{pmatrix} 1 \\ v \end{pmatrix} \le 0 \quad \forall i \in I(x, t). $$
		Expanding the inner product yields:
		$$ \frac{\partial g_i}{\partial t}(x, t) \cdot 1 + \nabla_x g_i(x, t) \cdot v \le 0. $$
		Since this inequality must hold for all possible velocity vectors $v \in f(x, t)$ permitted by the differential inclusion, we take the supremum over the set $f(x, t)$, strictly recovering \eqref{eq:C1_Condition}. This proves both necessity and sufficiency.
	\end{proof}
	
	\begin{rem}
		When $K(t)=K$ is {time-invariant}, $\partial g_i/\partial t\equiv0$ and
		condition~\eqref{eq:C1_Condition} reduces to
		\[
		\sup_{v\in f(x)}\nabla_x g_i(x)\cdot v\;\le\;0
		\qquad\forall\,x\in\partial K,\;\forall\,i\in I(x),
		\]
		which is precisely the classical algebraic Nagumo condition for differential
		inclusions~\cite[Thm.~5.3.4]{Aubin1991Viability}. 
	\end{rem}
	
	\begin{rem}
    For a single constraint ($m=1$) and an ODE $\dot{x}=f(x,u,t)$, \eqref{eq:C1_Condition} recovers the boundary form of the time-varying Control Barrier Function (CBF) $\dot{g}(x,u,t) \le \alpha(-g(x,t))$ proposed in \cite{ames2019control}. However, Theorem~\ref{thm:AlgebraicBarrier} provides three distinct improvements: 
    \begin{enumerate}
        \item  It establishes necessity and sufficiency, whereas CBFs are only sufficient  in the  nonsmooth case.
        \item It replaces the transversality condition (LICQ) with the weaker ACQ, allowing verification even where $\nabla_x g$ vanishes (e.g., at degenerate corners).
    \end{enumerate}
    Our framework thus unifies and generalizes CBF theory to multi-constraint, non-smooth environments under minimal regularity assumptions.
\end{rem}
    
		\subsection{Temporal Invariance of Time-Varying Polytopes for LTI Systems} \label{sec:problem_formulation_lti}
	
	Consider the continuous-time LTI system:
	\begin{equation} \label{eq:LinearSystemLTI}
		\dot{x}(t) = Ax(t), \qquad x(t_0) = x_0 \in \mathbb{R}^n,
	\end{equation}
	where $A \in \mathbb{R}^{n \times n}$ is constant.
	\begin{definition} \label{def:Polytope}
		Let $H: \mathbb{R}_{\ge 0} \to \mathbb{R}^{m \times n}$ and $h: \mathbb{R}_{\ge 0} \to \mathbb{R}^m$ be continuously differentiable. The {time-varying polytope} $K(t)$ is:
		\begin{equation}\label{eq:linearK(t)}
			K(t) := \{ x \in \mathbb{R}^n \mid H(t)x \le h(t) \} = \bigcap_{i=1}^m \{ x : H_i(t)x \le h_i(t) \},
		\end{equation}
		where $H_i(t)$ is the $i$-th row of $H(t)$ and $h_i(t)$ is the $i$-th element of $h(t)$.
	\end{definition}
	
	For index $i \in \{1, \dots, m\}$, the {$i$-th active face} at time $t$ is:
	\begin{equation}
		\mathcal{F}_i(t) := \{ x \in K(t) \mid H_i(t)x = h_i(t) \},
	\end{equation}
	a convex polytope of dimension at most $n-1$, possibly empty.
	
	For the deterministic LTI system \eqref{eq:LinearSystemLTI}, $f(x,t) = \{Ax\}$ is a singleton, so the supremum in \eqref{eq:C1_Condition} trivially reduces to a direct evaluation. The following theorem reduces the continuous-face verification to finitely many vertex checks.
	
	\begin{theorem}\label{thm:LinearVertexLTI}
		Consider the LTI system \eqref{eq:LinearSystemLTI} and the time-varying polytope $K(t)$ from Definition~\ref{def:Polytope}. The set $K(t)$ is temporally forward invariant if and only if, for every $i \in \{1, \dots, m\}$ and every vertex $v \in \text{vert}(\mathcal{F}_i(t))$:
		\begin{equation} \label{eq:VertexConditionLTI}
			\left( \dot{H}_i(t) + H_i(t)A \right) v \le \dot{h}_i(t).
		\end{equation}
	\end{theorem}
	
	\begin{proof}
By Theorem~\ref{thm:AlgebraicBarrier}, since the dynamics are single-valued $f(x,t)=Ax$, temporal forward invariance of $K(t)$ is equivalent to
\begin{equation}
\forall t \ge 0,\;\forall x \in \mathcal{F}_i(t),\;\forall j \in I(x,t):\quad \Phi_j(x,t) \le 0,
\end{equation}
where $\Phi_j(x,t) := \dot{H}_j(t)x - \dot{h}_j(t) + H_j(t)Ax$ and $I(x,t)$ is the set of active constraints at $x$.
By definition, for all $x \in \mathcal{F}_i(t)$, we have $i \in I(x,t)$, hence the invariance condition implies
\begin{equation}
\forall t \ge 0,\;\forall x \in \mathcal{F}_i(t):\quad \Phi_i(x,t) \le 0.
\end{equation}
For fixed $t$, the map $x \mapsto \Phi_i(x,t)$ is affine. Since $\mathcal{F}_i(t)$ is a  compact polytope\footnote{possibly lower-dimensional}, it follows that
\begin{equation}
\sup_{x \in \mathcal{F}_i(t)} \Phi_i(x,t)
=
\max_{v \in \mathrm{vert}(\mathcal{F}_i(t))} \Phi_i(v,t).
\end{equation}
\begin{itemize}
    \item[\textit{($\Leftarrow$):}] If \eqref{eq:VertexConditionLTI} holds for all vertices, then $\sup_{x \in \mathcal{F}_i(t)} \Phi_i(x,t) \le 0$ which implies $\Phi_i(x,t) \le 0$ for all $x \in \mathcal{F}_i(t)$. Since every $x \in K(t)$ satisfies $x \in \mathcal{F}_i(t)$ for all $i \in I(x,t)$, we obtain
\[
\Phi_j(x,t) \le 0 \quad \forall j \in I(x,t).
\]
Thus, the condition of Theorem~\ref{thm:AlgebraicBarrier} holds, and $K(t)$ is temporally forward invariant.
\item  [\textit{($\Rightarrow$):}] Conversely, if $K(t)$ is temporally forward invariant, then for all $t \ge 0$, all $x \in K(t)$, and all $j \in I(x,t)$,$\Phi_j(x,t) \le 0$.
In particular, for any $i$ and any vertex $v \in \mathrm{vert}(\mathcal{F}_i(t)) \subset K(t)$, we have $i \in I(v,t)$, hence $\Phi_i(v,t) \le 0$.
This proves the vertex condition.
\end{itemize}
	\end{proof}
	\subsection{Algebraic Characterization via Farkas' Lemma}
	The vertex-based condition in Theorem~\ref{thm:LinearVertexLTI} is exponentially complex in the worst case. The following proposition eliminates vertex enumeration entirely by recasting set-containment as an algebraic matrix condition via Farkas' Lemma.
	
	\begin{proposition}\label{prop:FarkasLTI}
		Consider the LTI system \eqref{eq:LinearSystemLTI} and the time-varying polytope \eqref{eq:linearK(t)}. The set $K(t)$ is temporally forward invariant if and only if, for every $t \ge 0$, there exists $\Lambda(t) \in \mathbb{R}^{m \times m}$ satisfying:
		\begin{subequations} \label{eq:FarkasConditions}
			\begin{align}
				\Lambda_{ij}(t) &\ge 0, \quad \forall i \neq j, \label{eq:MetzlerCondition} \\
				\dot{H}(t) + H(t)A &= \Lambda(t) H(t), \label{eq:MatrixEquality} \\
				\Lambda(t) h(t) &\le \dot{h}(t), \label{eq:VectorInequality}
			\end{align}
		\end{subequations}
		where $\Lambda(t)$ is required to be a Metzler matrix.
	\end{proposition}

	\begin{proof}
		By Theorem~\ref{thm:LinearVertexLTI}, invariance is equivalent to: for every $i \in \{1,\dots,m\}$, the linear functional $x \mapsto (\dot{H}_i(t) + H_i(t)A)x - \dot{h}_i(t)$ is non-positive on the active face
		\[
		\mathcal{F}_i(t) = \{x : H(t)x \le h(t),\; H_i(t)x = h_i(t)\}.
		\]
		This is a linear implication of the form: $H(t)x \le h(t)$ and $H_i(t)x = h_i(t)$ $\Rightarrow$ $c_i^\top x \le d_i$, where $c_i^\top = \dot{H}_i(t) + H_i(t)A$ and $d_i = \dot{h}_i(t)$. By the Affine Form of Farkas' Lemma \cite[Section~7.6]{schrijver1986theory}, this implication holds if and only if there exist non-negative multipliers $\lambda_{ij}(t) \ge 0$ for $j \neq i$ and an unconstrained multiplier $\lambda_{ii}(t) \in \mathbb{R}$, such that:
		\[
		c_i^\top = \sum_{j=1}^m \lambda_{ij}(t) H_j(t) \quad \text{and} \quad \sum_{j=1}^m \lambda_{ij}(t) h_j(t) \le \dot{h}_i(t).
		\]
		Stacking over $i = 1, \dots, m$ and defining $\Lambda(t)_{ij} := \lambda_{ij}(t)$ recovers exactly \eqref{eq:FarkasConditions}. The non-negativity of off-diagonal entries constitutes the Metzler condition \eqref{eq:MetzlerCondition}.
	\end{proof}
	
	\begin{remark}
	Proposition~\ref{prop:FarkasLTI} recasts temporal invariance verification in the algebraic space of $m \times m$ Metzler matrices rather than the geometric space $\mathbb{R}^n$. For fixed $t$, the feasibility of $\Lambda(t)$ satisfying \eqref{eq:FarkasConditions} is a Linear Program in $m^2$ variables, solvable in polynomial time. Critically, the problem size is independent of the state dimension $n$ and of the number of vertices, which can grow as $\Omega(2^n)$ for general polytopes.
	\end{remark}
	
	\subsection{Robust Discrete-Time Verification via Lipschitz Bounds}
	\label{sec:robust_disc}
	
	\begin{assumption}\label{assump:Lipschitz}
		Over the horizon $[0, \tau]$:
		\begin{itemize}
			\item $K(t)$ is uniformly bounded: there exists $R > 0$ such that $K(t) \subset \mathcal{B}_R$ for all $t \in [0,\tau]$.
			\item The trajectory functions $H(\cdot)$ and $h(\cdot)$ are twice continuously differentiable ($\mathcal{C}^2$) on $[0,\tau]$.
		\end{itemize}
	\end{assumption}
	
Under Assumption~\ref{assump:Lipschitz}, the evaluation functional $\Phi_i(x,t) := (\dot{H}_i(t) + H_i(t)A)x - \dot{h}_i(t)$ is uniformly Lipschitz continuous in $t$ over $\mathcal{B}_R \times [0,\tau]$. Indeed, since $H \in \mathcal{C}^2$ and $h \in \mathcal{C}^2$, the map $t \mapsto (\ddot{H}_i(t) + \dot{H}_i(t)A)x - \ddot{h}_i(t)$ is continuous on the compact set $\mathcal{B}_R \times [0,\tau]$ and hence bounded. This gives a global Lipschitz constant
	\begin{equation} \label{eq:LipschitzBound}
		L := \sup_{x \in \mathcal{B}_R,\, t \in [0,\tau],\, i} \left\| \frac{d}{dt}\bigl[(\dot{H}_i(t) + H_i(t)A)x - \dot{h}_i(t)\bigr] \right\| < \infty,
	\end{equation}
	
	so that $|\Phi_i(x, t_a) - \Phi_i(x, t_b)| \le L|t_a - t_b|$ for all $x \in \mathcal{B}_R$ and $t_a, t_b \in [0,\tau]$.

	\begin{proposition}\label{prop:RobustGridFarkas}
		Consider the LTI system \eqref{eq:LinearSystemLTI} and the time-varying polytope $K(t)$ from Definition~\ref{def:Polytope}. Suppose Assumption~\ref{assump:Lipschitz} holds. Let $\mathcal{T} = \{t_0, t_1, \dots, t_N\}$ be a finite grid on $[0,\tau]$ with maximum step size $\Delta t = \max_k(t_{k+1} - t_k)$ and Lipschitz constant $L$ as in \eqref{eq:LipschitzBound}. The set $K(t)$ is temporally forward invariant for all $t \in [0,\tau]$ if, at every $t_k \in \mathcal{T}$, there exists $\Lambda(t_k) \in \mathbb{R}^{m\times m}$ satisfying:
		\begin{subequations} \label{eq:RobustFarkasLP}
			\begin{align}
				\Lambda_{ij}(t_k) &\ge 0, \quad \forall i \neq j, \label{eq:RobustMetzler} \\
				\dot{H}(t_k) + H(t_k)A &= \Lambda(t_k) H(t_k), \label{eq:RobustMatrixEq} \\
				\Lambda(t_k) h(t_k) &\le \dot{h}(t_k) - L\,\Delta t\cdot \mathbf{1}, \label{eq:RobustVectorIneq}
			\end{align}
		\end{subequations}
		where $\mathbf{1} \in \mathbb{R}^m$ is the all-ones vector.
	\end{proposition}
	
	\begin{proof}
		Suppose $\Lambda(t_k)$ satisfies \eqref{eq:RobustFarkasLP} at grid point $t_k$. By applying the Farkas implication backward from \eqref{eq:RobustMetzler}--\eqref{eq:RobustVectorIneq} (reversing the argument of Proposition~\ref{prop:FarkasLTI} with the tightened right-hand side), we obtain:
		\begin{equation}
			\max_{x \in \mathcal{F}_i(t_k)} \Phi_i(x, t_k) \le -L\Delta t, \quad \forall i \in \{1, \dots, m\}.
		\end{equation}
		Now let $t \in (t_k, t_{k+1})$ and $x \in \mathcal{F}_i(t)$. Note that $x$ need not belong to $\mathcal{F}_i(t_k)$; however, since $K(t) \subseteq \mathcal{B}_R$ for all $t$ by Assumption~\ref{assump:Lipschitz}, we have $x \in \mathcal{B}_R$, and the Lipschitz bound \eqref{eq:LipschitzBound} applies point-wise in $x$. Therefore:
		\begin{align*}
			\Phi_i(x, t) &\le \Phi_i(x, t_k) + L|t - t_k| \\
			&\le \max_{y \in \mathcal{B}_R} \Phi_i(y, t_k) + L\Delta t \\
			&\le \!\!\!\max_{y \in \mathcal{F}_i(t_k)} \!\!\!\Phi_i(y, t_k) + L\Delta t \quad (\text{since } \mathcal{F}_i(t_k)\!\! \subseteq\! K(t_k) \! \subseteq \!\mathcal{B}_R \!) \\
			&\le (-L\Delta t) + L\Delta t = 0.
		\end{align*}
		Hence $\Phi_i(x,t) \le 0$ for all $x \in \mathcal{F}_i(t)$ and all $t \in [0,\tau]$. By Theorem~\ref{thm:LinearVertexLTI}, $K(t)$ is temporally forward invariant.
	\end{proof}
	
	\begin{remark}
		Proposition~\ref{prop:RobustGridFarkas} reduces the infinite-dimensional verification problem to exactly $N+1$ decoupled Linear Programs, each with $m^2$ decision variables, independent of the state dimension $n$ and the number of vertices of $K(t)$. Using interior-point methods, each LP is solved in $\mathcal{O}(m^{3.5})$ time, yielding a total worst-case complexity of $\mathcal{O}(N \cdot m^{3.5})$.
	\end{remark}

\vspace{-5pt}
	\section{Case studies}    
	\label{sec:case_studies}
	
	\subsection{Temporal Forward Invariance for a 2D Oscillatory System}
	
	In this subsection, we apply Theorem \ref{thm:LinearVertexLTI} to a stiff linear harmonic oscillator. We demonstrate that a time-varying polytopic set $K(t)$, constructed by transporting an initial geometry along the system flow, satisfies the invariance conditions identically.
	
	\subsubsection{System Dynamics and Set Definition}
	
	We consider a continuous-time LTI system representing an asymmetric harmonic oscillator. The dynamics are given by \eqref{eq:LinearSystemLTI} with the matrix:
	\begin{equation} \label{eq:ExampleSystem}
		A = \begin{bmatrix} 
			0 & -\omega_1 \\ 
			\omega_2 & 0 
		\end{bmatrix} = \begin{bmatrix} 
			0 & -1 \\ 
			100 & 0 
		\end{bmatrix}.
	\end{equation}
	The eigenvalues of $A$ are $\lambda_{1,2} = \pm j10$, indicating stable oscillatory behavior with a period $T = \pi/5 \approx 0.628$s. The asymmetry in the off-diagonal terms ($\omega_2 \gg \omega_1$) results in trajectories that are elongated ellipses, aligned with the $x_2$-axis.
	
	The time-varying set $K(t)$ is defined as the image of a static initial box $K_0 = [-1, 1] \times [-1, 1]$ under the system's state transition map. Mathematically, the initial set is defined by the inequality $H_0 x \le h_0$, where $H_0 \in \mathbb{R}^{4 \times 2}$ contains the standard basis vectors and their negatives, and $h_0 = \mathbf{1}_4$. For $t > 0$, the set is defined by the transport equation:
	\begin{equation} \label{eq:ExampleSet}
		K(t) = \{ x \in \mathbb{R}^2 \mid H(t)x \le h(t) \}, \quad \text{with } \begin{cases} H(t) = H_0 e^{-At} \\ h(t) = h_0 \end{cases}.
	\end{equation}
	
	\subsubsection{Analytical Verification of temporal forward Invariance}
	
	To verify the invariance of $K(t)$, we check the vertex condition from Theorem \ref{thm:LinearVertexLTI}. The condition requires that for every active vertex $v(t)$ of the face $\mathcal{F}_i(t)$, the inequality $(\dot{H}_i(t) + H_i(t)A)v(t) \le \dot{h}_i(t)$ holds.
	
\begin{figure}
		\centering
		\includegraphics[width=1\linewidth]{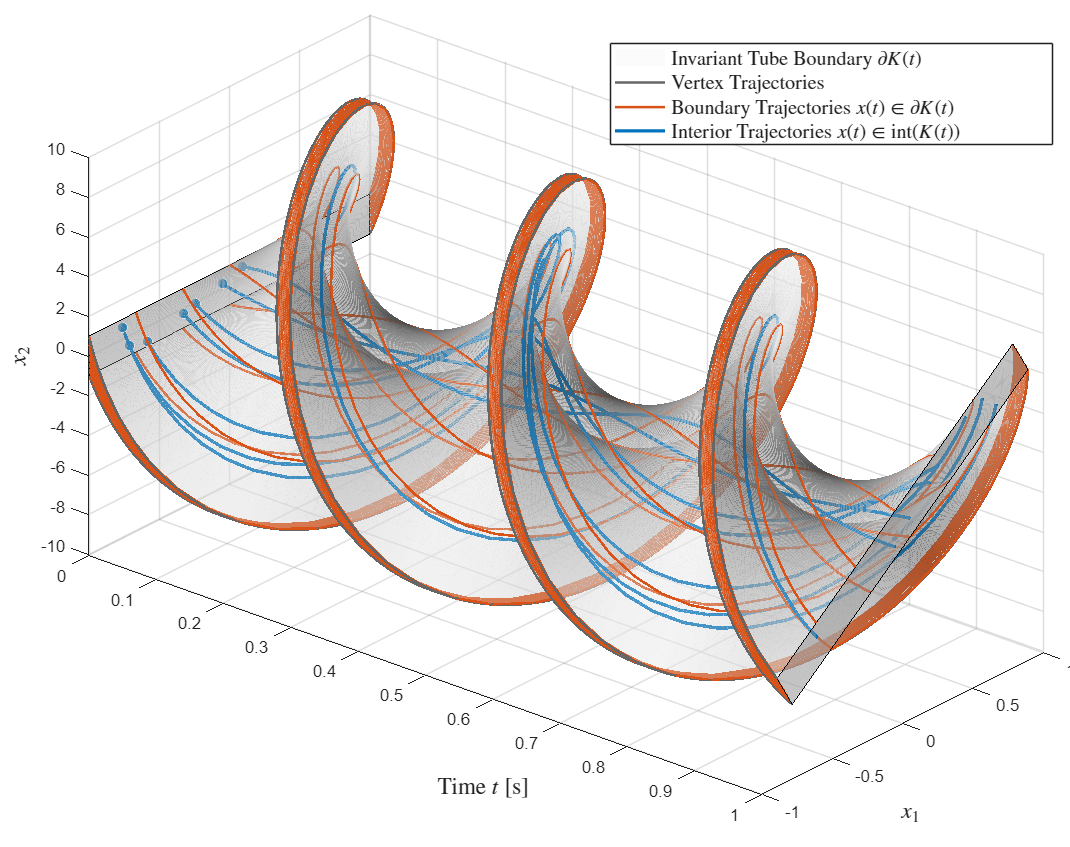}
		\caption{Spatiotemporal graph of the invariant polytope $K(t)$ containing the asymmetric oscillator trajectories}
		\label{fig:1}
	\end{figure}

	First, we compute the time derivative of the constraint matrix $H(t)$. Using the property $\frac{d}{dt} e^{At} = A e^{At} = e^{At} A$ (see \cite{strang2006linear}), we obtain:
	\begin{align*}
		\dot{H}(t) &= \frac{d}{dt} \left( H_0 e^{-At} \right) \\
		&= H_0 \left( -A e^{-At} \right) \\
		&= - \left( H_0 e^{-At} \right) A \\
		&= -H(t)A.
	\end{align*}
	Since the right-hand side vector $h(t)$ is constant, $\dot{h}(t) = 0$. Substituting these results into the evaluation functional $\Phi_i(v, t)$ for any vertex $v$:
	\begin{equation}
		\Phi_i(v, t) = \left( \underbrace{-H_i(t)A}_{\dot{H}_i(t)} + H_i(t)A \right) v - \underbrace{0}_{\dot{h}_i(t)} = 0.
	\end{equation}
	Condition  \eqref{eq:VertexConditionLTI} holds with equality $0$. This confirms that $K(t)$ is exactly invariant. Thus, the resultant spatiotemporal graph of the time varying polytope $K(t)$ and  the asymmetric oscillator trajectories are illustrated in Fig. \ref{fig:1}.
	
	\subsection{Verification of a grid-forming inverter}
	
	To demonstrate the validity of Theorem \ref{thm:LinearVertexLTI}, we consider the transient safety of an islanded microgrid inverter. In this scenario, the objective is to ensure that frequency deviations following a load disturbance remain within a time-varying recovery funnel that contracts toward a strict steady-state tolerance.
	
	\subsubsection{Microgrid System Dynamics}
	We adopt a linearized 2D model of a Virtual Synchronous Generator (VSG) based on the swing equation. The state vector is defined as $x = [\delta, \omega]^T$, where $\delta$ represents the power angle and $\omega$ is the frequency deviation from the nominal $50/60$ Hz. The dynamics are governed by:
	\begin{equation}
		\begin{bmatrix} \dot{\delta} \\ \dot{\omega} \end{bmatrix} = 
		\begin{bmatrix} 
			0 & 1 \\ 
			-\frac{K_p}{M} & -\frac{D}{M} 
		\end{bmatrix} 
		\begin{bmatrix} \delta \\ \omega \end{bmatrix}
	\end{equation}
	Following standard parameters for low-inertia microgrids \cite{Bevrani2014}, we set the virtual inertia $M = 0.1 \text{ pu}\cdot\text{s}$, damping $D = 0.5 \text{ pu}$, and synchronizing power $K_p = 2.0 \text{ pu}$.
	
	We define a time-varying polytopic set $K(t) = \{ x \in \mathbb{R}^2 \mid |x_i| \le h_i(t) \}$. The boundary for frequency deviation, $h_\omega(t)$, is designed as a "Frequency Recovery Funnel" that contracts exponentially to satisfy grid code standards (e.g., IEEE 1547 \cite{IEEE1547_2018}):
	\begin{equation}
		h_\omega(t) = h_{\omega, \infty} + (h_{\omega, 0} - h_{\omega, \infty})e^{-\lambda t}
	\end{equation}
	where $h_{\omega, 0} = 1.0$ rad/s is the initial fault tolerance, $h_{\omega, \infty} = 0.1$ rad/s is the steady-state limit, and $\lambda$ is the required contraction rate. To maintain set invariance, the angle boundary $h_\delta(t)$ is allowed to expand initially to accommodate phase drift: $h_\delta(t) = 0.5 + \int_0^t h_\omega(\tau)d\tau$.

	\subsubsection{Analytical Verification of temporal forward Invariance}	
\begin{figure}[htbp]
		\centering
		\includegraphics[width=0.5\textwidth]{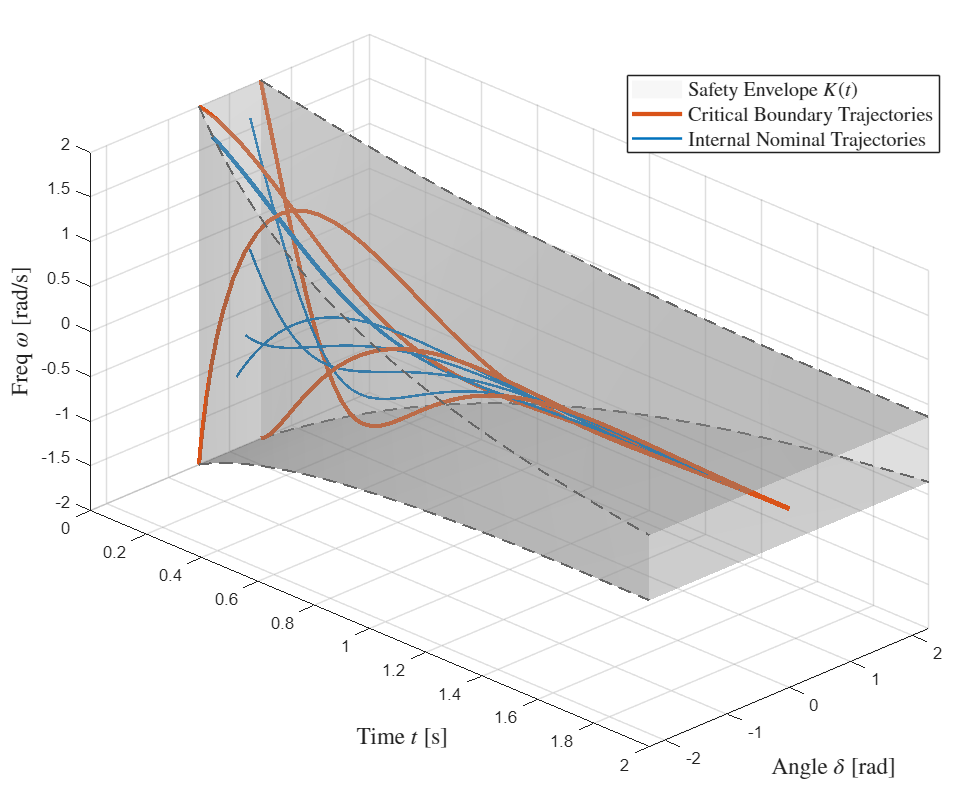}
		\caption{Spatiotemporal safety tube $K(t)$ for the VSG model. The trajectories remain strictly within the contracting frequency bounds as the condition \eqref{eq:VertexConditionLTI} is satisfied.}
		\label{fig:microgrid_funnel}
	\end{figure}

	According to Theorem \ref{thm:LinearVertexLTI}, the set $K(t)$ is invariant if $H_i A v \le \dot{h}_i(t)$ for all vertices $v$ on the $i$-th face. For the frequency upper bound ($H_\omega = [0, 1]$), the condition is evaluated at the critical vertex $v^* = [-h_\delta(t), h_\omega(t)]^T$, where the restoring force $-\frac{K_p}{M}\delta$ is most antagonistic to the boundary contraction:
	\begin{equation}
		\underbrace{-\frac{K_p}{M}(-h_\delta(t)) - \frac{D}{M}h_\omega(t)}_{\text{System Acceleration}} \le \underbrace{-\lambda(h_{\omega, 0} - h_{\omega, \infty})e^{-\lambda t}}_{\text{Envelope Contraction Velocity}}
	\end{equation}
	Physically, the term $\frac{D}{M}$ represents the mechanical decay rate provided by the inverter's damping logic. Theorem \ref{thm:LinearVertexLTI} reveals that the safety tube remains valid as long as the damping-to-inertia ratio is sufficiently high to outrun the required contraction rate $\lambda$. If $\lambda$ is chosen such that the envelope contraction velocity becomes more negative than the system acceleration, the physical system cannot restore frequency fast enough to stay within the tightening bounds, resulting in a safety violation. As shown in Fig. \ref{fig:microgrid_funnel}, a design with $\lambda = 1.5$ satisfies the vertex condition, ensuring all trajectories starting within $K(0)$ remain safe for all $t > 0$.
	
    \subsubsection{Algebraic Verification via Farkas' Lemma}

According to Proposition~\ref{prop:FarkasLTI}, the set $K(t)$ is temporally forward invariant if there exists a Metzler matrix $\Lambda(t)$ such that
\begin{equation}
\begin{array}{rl}
     \dot{H}(t) + H A &   = \Lambda(t) H\\
     \Lambda(t) h(t) &  \le \dot{h}(t)
\end{array}
\end{equation}

For the box constraints, we consider $
H = \begin{bmatrix}
1 & 0\\
-1 & 0\\
0 & 1\\
0 & -1
\end{bmatrix}.
$

Solving the linear system $H A = \Lambda H$ yields a family of solutions parameterized by nonnegative off-diagonal entries of $\Lambda$. Selecting a feasible Metzler matrix (e.g., by setting a subset of coefficients to zero and enforcing nonnegativity constraints), the inequality $\Lambda(t) h(t) \le \dot{h}(t)$ reduces to a set of linear inequalities.

In particular, one recovers the dominant condition:
\begin{equation}
\frac{K_p}{M} h_\delta(t) - \frac{D}{M} h_\omega(t)
\le
\dot{h}_\omega(t),
\end{equation}
which coincides with the vertex-based condition derived above. 
	\section{Conclusion }
	
	This paper has established a rigorous geometric and algebraic framework for the temporal forward invariance of time-varying sets. By lifting the system dynamics into an augmented spatiotemporal graph, we proved that the Nagumo-type contingent-cone condition is both necessary and sufficient when restricted to the boundary of the graph. Under the Abadie Constraint Qualification, this geometric condition was translated into an algebraic barrier condition that generalizes time-varying CBFs to multi-constraint, non-smooth domains---recovering the classical CBF boundary condition as a special case under LICQ. For LTI systems with time-varying polytopic constraints, we derived a vertex condition and, via Farkas' Lemma, eliminated vertex enumeration entirely, reducing the problem to Metzler-constrained Linear Programs. A Lipschitz-bounded discretization scheme further reduced the continuous-time horizon to a finite grid of decoupled LPs with polynomial complexity $\mathcal{O}(N \cdot m^{3.5})$.
	 
     Future directions include: (i)~extending the ACQ-based algebraic characterization to nonlinear time-varying sets beyond the polytopic case, leveraging recent advances in high-order CBFs \cite{xiao2022sufficient}; (ii)~incorporating disturbance inputs and robust invariance via the future-focused CBF techniques of \cite{black2023safety}; and (iii)~addressing Signal Temporal Logic specifications through the polytopic invariance framework, building on the CBF-for-STL approaches of \cite{lindemann2019cbf,xiao2021hoclbf}.

	\bibliographystyle{ieeetr}
	\bibliography{biblio}

\end{document}